\RequirePackage{fix-cm}
\documentclass[smallextended]{svjour3}       
\smartqed  
\usepackage{graphicx}
\usepackage{epsfig,latexsym,amssymb,amsmath}
\begin{document}

\title{Thermodynamic Products for Sen Black Hole      
}


\author{Parthapratim Pradhan}
\institute{\at Department of Physics\\
           Vivekananda Satavarshiki Mahavidyalaya \\
           (Affiliated to Vidyasagar University)\\
            Manikpara, Jhargram, West Midnapur\\
            West Bengal-721513, India. \\
            \email{pppradhan77@gmail.com}}
\date{Received: date / Revised version: date}

\maketitle

\begin{abstract}
We investigate the properties of  inner and outer horizon thermodynamics of Sen  black hole(BH) both in
\emph{Einstein frame(EF)} and \emph{string frame(SF)}. We also compute area(or entropy) product, area(or entropy) 
sum of the said BH  in EF as well as SF. In the EF, we observe that the area(or entropy) product is \emph{ universal}, 
whereas area (or entropy) sum is \emph{not}  universal. On the other hand,  in the SF, area(or entropy) product and 
area(or entropy) sum don't have any universal behaviour because they  all  are depends on ADM(Arnowitt-Deser-Misner) mass
parameter. We also verify that the \emph{first law} is satisfied at the Cauchy horizon(CH) as well as event horizon(EH). 
In addition, we also compute other thermodynamic products and sums in the EF as well as in the SF. We  
further compute the \emph{Smarr mass formula} and \emph{Christodoulou's irreducible mass formula} for Sen BH. Moreover, we 
compute the area bound and entropy bound for both the horizons. The upper area bound for EH is actually the Penrose like 
inequality, which is the first geometric inequality in BHs. Furthermore, we compute the central charges of the left and 
right moving sectors of the dual CFT in Sen/CFT correspondence using thermodynamic relations. These thermodynamic 
relations on the multi-horizons give us further understanding the microscopic nature of BH 
entropy(both interior and exterior).
\end{abstract}

\section{Introduction}
In an un-quantized(classical) general relativity theory any BH in thermal equilibrium has an entropy and a
temperature. Now it is well known by fact that the entropy is proportional to the area of the event horizon(EH)
i.e. \cite{bk72,bk73,bk74,bcw73}
\begin{eqnarray}
{\cal S}_{+} &=& \frac{{\cal A}_{+}}{4} ~. \label{splus}
\end{eqnarray}
where, ${\cal S}_{+}$ is the Bekenstein-Hawking entropy (in units in which $G=\hbar=c=k=1$) and ${\cal A}_{+}$ is
the area of the EH(${\mathcal H}^{+}$).
Now this temperature is proportional to the surface gravity of
the ${\mathcal H}^{+}$ i.e.
\begin{eqnarray}
T_{+} &=& \frac{\kappa_{+}}{2 \pi}  ~. \label{tplus}
\end{eqnarray}
where $T_{+}$ is the Hawking temperature computed at the ${\mathcal H}^{+}$ and 
$\kappa_{+}$ denotes the surface gravity of the BH computed at the ${\mathcal H}^{+}$.

In terms of these quantities, the first law of BH thermodynamics could be expressed as
\begin{eqnarray}
dM &=& \frac{{\kappa}_{+}} {8\pi} d{\cal A}_{+} + \Omega_{+} dJ +\Phi_{+}dQ ~. \label{dm1}
\end{eqnarray}
It can be seen that $\frac{{\kappa}_{+}} {8\pi}$ is analogous to the temperature of ${\cal H}^{+}$
in the same way that ${\cal A}_{+}$ is analogous to entropy. It should be noted that
$\frac{{\kappa}_{+}} {8\pi}$ and ${\cal A}_{+}$ are  distinct from the temperature and entropy of the BH.
and
\begin{eqnarray}
\Omega_{+} &=& \frac{4\pi J}{ M{\cal A}_{+}}=\frac{\partial  M}{\partial J} \\
\Phi_{+} &=& \frac{1}{M} \left(\frac{Q}{2}+\frac{2\pi Q^3}{{\cal A}_{+}}  \right) =
\frac{\partial M}{\partial Q} ~. \label{invar1}
\end{eqnarray}
The above relations are computed on the EH only.

It is now well known fact that certain BH has inner horizon or Cauchy horizon(CH) inside
the EH or outer horizon. Naturally, the question should be arises whether similar relations
do exist in case of CH? It is now well established that the above relations do
hold for CH (${\mathcal H}^{-}$) as well as EH. Therefore one may write the inner entropy of
the BH which is proportional to the area of the inner horizon:
\begin{eqnarray}
{\cal S}_{-} &=& \frac{{\cal A}_{-}}{4}  ~. \label{sminus}
\end{eqnarray}
Analogously, the inner Hawking temperature should be calculated via the inner surface gravity
of the BH:
 \begin{eqnarray}
T_{-} &=& \frac{\kappa_{-}}{2 \pi}  ~. \label{tminus}
\end{eqnarray}
Using the above inner properties of the BH, we can write the inner first law of BH
thermodynamics 
\begin{eqnarray}
dM &=& -\frac{{\kappa}_{-}} {8\pi} d{\cal A}_{-} + \Omega_{-} dJ +\Phi_{-}dQ ~. \label{dm2}
\end{eqnarray}
where,
\begin{eqnarray}
\Omega_{-} &=& \frac{4\pi J}{ M{\cal A}_{-}}=\frac{\partial  M}{\partial J} \\
\Phi_{-} &=& \frac{1}{M} \left(\frac{Q}{2}+\frac{2\pi Q^3}{{\cal A}_{-}}  \right) =
\frac{\partial M}{\partial Q} ~. \label{invar2}
\end{eqnarray}

Similarly, the second law is also valid for outer horizon\cite{bcw73} as well as inner
horizon which states that
 \begin{eqnarray}
d{\cal A}_{\pm} & \geq & 0  ~. \label{2nd}
\end{eqnarray}

It has been suggested that every regular axi-symmetric and stationary space-time of
Einstein-Maxwell gravity with surrounding matter has a regular CH inside the EH
if and only if  both angular momentum $J$ and charge $Q$ do not vanish. Then
the product of the  area $A_{\pm}$ of the horizons ${\cal H}^\pm$ for KN class of family
could be  expressed as by the relation \cite{ansorg09}:  (see also \cite{visser12,pp14})
\begin{eqnarray}
{\cal A}_{+} {\cal A}_{-} &=& (8\pi)^2\left(J^2+\frac{Q^4}{4}\right) ~.\label{proarKN}
\end{eqnarray}
which is remarkably independent of the ADM mass $(M)$ of the space-time. In the limit $Q=0$, one
obtains the area product formula for Kerr BH\cite{cr79}. 

Again in string theory and $M$-theory, the product of Killing horizon areas for certain multi-horizon
BHs are also independent of the ADM mass.  For asymptotically flat BPS
(Bogomol'ni-Prasad-Sommerfield) BHs in four and higher dimensions, the quantization
rule  becomes\cite{mcdy96,mcflb97,mcfl97}: 
${\cal A}_{\pm}  =8\pi {\ell _{pl}}^2 \left(\sqrt{N_{1}}\pm\sqrt{N_{2}} \right)$ or
\begin{eqnarray}
{\cal A}_{+} {\cal A}_{-} &=& \left(8\pi {\ell _{pl}}^2\right)^2 N , \,\,\, N\in {\mathbb{N}},
N_{1}\in {\mathbb{N}}, N_{2} \in {\mathbb{N}}  ~.\label{ppl}
\end{eqnarray}
where $\ell _{pl}$ is the Planck length,  $N_{1}$ and $N_{2}$ are  integers for super-symmetric extremal
BHs\cite{mcfl97,mcflb97,mcgw11,pope14,castro12,det12,val13,chen12,hl15}.

However, it has been well known fact that CH is a infinite blue-shift region in contrast with
EH(infinite red-shift region). It is also true that CH is a highly unstable due to the exterior
perturbation\cite{sch}. Thus there has been indication towards the relevance of \emph{BH CH}
in comparison with EH.

Thus in this work, we wish to examine the above mentioned thermodynamical feature of
the rotating charged BHs in heterotic string theory\cite{as92}\footnote{The Sen's 1992\cite{as92} 
solutions were later generalized by Sen 1995\cite{as94} and also Cvetic and Youm 1996\cite{mcdy96}. 
Since those times one standard class for many investigations of string theory BHs were the four 
charge solutions parametrized by four boost angles. The Sen 1992 solutions\cite{as92} correspond 
to the special case where three of the four boost angles were taken to vanish. The present manuscript 
is a special case, where three parameters are taken to vanish.}.  We have discussed both the
situations in``Einstein frame'' as well as in ``String frame''. The fact that string theory is the
leading candidate to unify gravity to other fundamental forces in nature. For this reason,
we have chosen the  low energy  heterotic string theoretical BH. The special characteristics
of this string BH is that they are qualitatively different from those BH that appear in
ordinary Einstein general theory of relativity\cite{as92,blaga01}. Most of these solutions are
characterized by one or more charges associated with Yang-Mills fields or the anti-symmetric
tensor gauge field. Furthermore, this low energy heterotic string BH carries a finite amount
of charge, angular momentum and magnetic dipole moment. It could be produced
by twisting method and starting from a rotating BH having no charge, i.e. the Kerr BH.
 So, sometimes it is called twisted Kerr BH or Kerr-Sen BH\cite{blaga01}.

We prove that in the Einstein frame, the area product formula and the entropy product formula 
are universal, whereas area sum and entropy sum are not universal. Whereas in the string frame, 
area product, entropy product, area sum and entropy sum formula don't have any universal nature 
because they  all  are depends on ADM mass parameter. We also observe that every BH thermodynamic 
quantities (e.g. area, entropy, temperature, surface gravity  etc.), other than the mass ($M$), 
the angular momentum ($J$) and the charge $(Q)$, can form a quadratic equation whose roots
are contained the three basic parameter $M,J,Q$. We further examine that the four laws of BH 
mechanics is satisfied at the inner horizon as well as EH. Moreover in Einstein frame, we compute 
the area bound and entropy bound for both the horizons. The upper area bound for event horizon is 
actually the Penrose like  inequality, which is the first geometric inequality in BHs\cite{peni}.

The paper is organized as follows. Sec. \ref{ksen} describes the properties of Sen BH in Einstein
frame and deals with various thermodynamic products. In this section, there are three subsections.
In first subsection \ref{smar},  we have discussed the Smarr formula
for Sen BH. In second subsection\ref{ruffini}, we have derived the Christodoulou-Ruffini  mass formula
for Sen BH. Finally in third subsection\ref{law}, we have discussed the four laws of BH thermodynamics.
In Sec. \ref{sen}, we computed various thermodynamic products for Sen BH in String frame. Finally,
in Sec. \ref{dis} we concluded our discussions.

\section{\label{ksen} Sen BH in Einstein Frame: }

An exact rotating charged BH solution in four dimension heterotic string theory represented
by the metric \cite{as92} in  Einstein frame
$$
ds^2 = -\left(1-\frac{2mr \cosh^2\alpha }{\rho^2}\right)dt^2-\frac{4amr \cosh^2 \alpha \sin^2\theta}{\rho^2}dt d\phi +
\frac{\rho^2}{\Delta}dr^2
$$
\begin{eqnarray}
+\rho^2 d\theta^2+ \frac{\Upsilon}{\rho^2} \sin^2\theta d\phi^2 ~\label{seneq}
\end{eqnarray}
where
\begin{eqnarray}
 \rho^2 &=& r^2+a^2\cos^2\theta+2mr \sinh^2\alpha \\
 \Delta &=& r^{2}-2mr+a^2 \\
 \Upsilon &=&( r^2+a^2+2mr \sinh^2\alpha)^2-\Delta a^2\sin^2\theta
\end{eqnarray}
The Maxwell field, dilaton, and anti-symmetric tensor are
\begin{eqnarray}
A &=& \frac{mr\sinh 2\alpha}{\sqrt{2}\rho^2}\left( dt-a \sin^2\theta d\phi\right) \\
e^{-2\phi} &=& \frac{\rho^2}{r^2+a^2 \cos^2\theta}\\
B_{t\phi} &=& \frac{2mar \sinh^2\alpha \sin^2\theta}{\rho^2}
\end{eqnarray}
The above metric describes a BH solution with mass $M$, charge $Q$, angular momentum $J$, and
magnetic dipole moment $\mu$ is given by
\begin{eqnarray}
M &=& \frac{m}{2}\left( 1+\cosh {2\alpha}\right)\\
Q &=&  \frac{m}{\sqrt{2}} \sinh 2\alpha\\
J &=&  \frac{ma}{2} \left( 1+\cosh 2\alpha\right)\\
\mu &=&  \frac{1}{\sqrt{2}} ma \sinh 2\alpha  ~\label{mqj}
\end{eqnarray}
Since we shall analyze various thermodynamic products of this BH, for this purpose it will
be more convenient to write $m$, $a$ and $\alpha$ in terms of the independent physical
parameters $M$, $J$ and $Q$ inverting the relations given in Eq. \ref{mqj}. Thus we find
\begin{eqnarray}
m &=& M-\frac{Q^2}{2M}\\
\sinh 2\alpha &=&  \frac{2\sqrt{2} QM}{(2M^2-Q^2)}\\
a &=&  \frac{J}{M}   ~\label{mqj1}
\end{eqnarray}
This is the well known Sen BH solution\cite{as92} which was discovered by Sen in 1992.

There are two horizons for Sen BH namely EH (${\cal H}^+$) or outer horizon
and CH (${\cal H}^-$) or inner horizon. There radius can be determined by
solving the following metric functions as
\begin{eqnarray}
\Delta \equiv \Delta (r) = r^{2}-\left(2M-\frac{Q^2}{M}\right)r+a^2=0  ~.\label{horizon}
\end{eqnarray}
which gives
\begin{eqnarray}
r_{\pm}&=& \left(M-\frac{Q^2}{2M}\right)\,\, \pm \sqrt{\left(M-\frac{Q^2}{2M}\right)^{2}-a^2}
~.\label{horradii}
\end{eqnarray}

Here $r_{+}$ is called event horizon(EH) and $r_{+}$ is called Cauchy horizon(CH). It may be
noted that $r_{+}> r_{-}$. Interestingly, the solution of the Eq. (\ref{horizon})
gives
\begin{eqnarray}
r_{+}+ r_{-}=2M-\frac{Q^2}{M} \mbox{and}\,\, r_{+} r_{-} &=& a^2 ~.\label{spsen}
\end{eqnarray}
This indicates that the sum and product of the horizon radii depends on the mass parameter.

From Eq.(\ref{horradii}) one can see that the horizon disappears unless
\begin{eqnarray}
a\leq\left(M-\frac{Q^2}{2M}\right).~\label{ieqs}
\end{eqnarray}
Thus the extremal limit of the Sen BH corresponds to
\begin{eqnarray}
a=\left(M-\frac{Q^2}{2M}\right).
\end{eqnarray}
and the horizon for extremal Sen BH is situated at
\begin{eqnarray}
r_{ex}=r_{+}=r_{-}=a=\left(M-\frac{Q^2}{2M}\right).
\end{eqnarray}

Now we would like to compute various thermodynamic quantities of the Sen BH.
The  area\cite{bk73,bk74} of both the horizon (${\cal H}^\pm$) in
Einstein frame (EF) is
\begin{eqnarray}
{\cal A}_{\pm} &=& \int^{2\pi}_0\int^\pi_0  \sqrt{g_{\theta\theta}g_{\phi\phi}}d\theta d\phi\\
               &=& 8\pi M \left[\left(M-\frac{Q^2}{2M}\right) \pm
                   \sqrt{\left(M-\frac{Q^2}{2M}\right)^{2}-a^2}\right] ~.\label{areasen}
\end{eqnarray}
The angular velocity of ${\cal H}^\pm$ computed at the horizon is given by
\begin{eqnarray}
{\Omega}_{\pm} &=& \frac{J}{2M^{2}\left[\left(M-\frac{Q^2}{2M}\right) \pm
\sqrt{\left(M-\frac{Q^2}{2M}\right)^{2}-a^2}\right]} ~. \label{omega}
\end{eqnarray}

The semi-classical Bekenstein-Hawking entropy of ${\cal H}^\pm$ reads
\begin{eqnarray}
{\cal S}_{\pm} &=& 
2\pi M \left[\left(M-\frac{Q^2}{2M}\right) \pm
\sqrt{\left(M-\frac{Q^2}{2M}\right)^{2}-a^2}\right] ~.\label{entsen}
\end{eqnarray}

The surface gravity\cite{as92} of ${\cal H}^\pm$ is given by
\begin{eqnarray}
{\kappa}_{\pm} &=& \pm \frac{\sqrt{(2M^2-Q^2)^2-4J^2}}{2M[(2M^2-Q^2)\pm \sqrt{(2M^2-Q^2)^2-4J^2}]} ~.\label{sgKN}
\\
\,\, \mbox{and} \nonumber\\
\,\,  \kappa_{+}> \kappa_{-} 
\end{eqnarray}
and the BH temperature or Hawking temperature of ${\cal H}^\pm$ reads as
\begin{eqnarray}
T_{\pm}&=& \pm \frac{\sqrt{(2M^2-Q^2)^2-4J^2}}{4\pi M[(2M^2-Q^2)\pm \sqrt{(2M^2-Q^2)^2-4J^2}]} ~.\label{tmKN}
\end{eqnarray}
It should be noted that $T_{+} > T_{-} $.

The Komar\cite{komar59} energy for  ${\cal H}^\pm$ is given by
\begin{eqnarray}
E_{\pm} &=& \pm \sqrt{(2M^2-Q^2)^2-4J^2} ~. \label{komeng}
\end{eqnarray}

Finally, the horizon Killing vector field may be defined for ${\cal H}^\pm$ is
\begin{eqnarray}
{\chi_{\pm}}^{a} &=& (\partial_{t})^a +\Omega_{\pm} (\partial_{\phi})^a~.\label{hkv}
\end{eqnarray}

Now we shall see that every BH thermodynamic quantities(e.g. area, entropy, temperature,
surface gravity  etc.), other than the mass ($M$), the angular momentum ($J$) and the charge $(Q)$, which
is also defined on ${\cal H}^{\pm}$ can form a quadratic equation of thermodynamic quantities like horizon
radii $(r_{\pm})$.

Firstly, we compute the ``product'' and ``sum'' of the inner horizon area and outer
horizon area as
\begin{eqnarray}
{\cal A}_{-}{\cal A}_{+}  &=& \left(8\pi J\right)^2 ~.\label{prarsen}
\end{eqnarray}
and
\begin{eqnarray}
{\cal A}_{-}+{\cal A}_{+}  &=& 8\pi\left(2M^2-Q^2 \right)~.\label{sumarsen}
\end{eqnarray}

Interestingly, the area sum and the area product might be satisfied the following
quadratic equation:
\begin{eqnarray}
{\cal A}^{2}-8\pi\left(2M^2-Q^2 \right){\cal A}+ \left(8\pi J\right)^2 &=& 0 ~.\label{quad1}
\end{eqnarray}
With the help of the above Eq. (\ref{prarsen}) and (\ref{sumarsen}), we can easily
see that the ``area product'' is universal, while the ``area sum'' is not universal
for Sen BH in Einstein frame because it depends on the BH mass or ADM mass parameter.
For completeness, we further compute the area minus and area division, which is given by
\begin{eqnarray}
{\cal A}_{\pm}- {\cal A}_{\mp} &=& 8\pi M T_{\pm}{\cal A}_{\pm}  ~.\label{area-}
\end{eqnarray}
and  
\begin{eqnarray}
\frac{{\cal A}_{+}}{{\cal A}_{-}} &=& \frac{r_{+}}{r_{-}}=\frac{\Omega_{-}}{\Omega_{+}}
=-\frac{T_{-}}{T_{+}} ~.\label{aread}
\end{eqnarray}

Again, the sum of area inverse is found to be 
\begin{eqnarray}
\frac{1}{{\cal A}_{+}}+\frac{1}{{\cal A}_{-}} &=& \frac{2 M^2-Q^2} {8 \pi J^2} ~.\label{arid}
\end{eqnarray}
and the minus of area inverse is computed to be 
\begin{eqnarray}
\frac{1}{{\cal A}_{\pm}}-\frac{1}{{\cal A}_{\mp}} &=& \mp \frac{\sqrt{(2M^2-Q^2)^2-4J^2}}{8\pi J^2}
~.\label{areid-}
\end{eqnarray}
It indicates that they all are mass dependent relations.

Likewise, the ``entropy product'' and ``entropy sum'' of ${\cal H}^\pm$
becomes:
\begin{eqnarray}
{\cal S}_{-} {\cal S}_{+}  &=& \left(2\pi J\right)^2 ~.\label{etpsen}
\end{eqnarray}
and
\begin{eqnarray}
{\cal S}_{-}+{\cal S}_{+}  &=& 2\pi\left(2M^2-Q^2 \right) ~.\label{entpsum}
\end{eqnarray}
The quadratic equation of entropy becomes
\begin{eqnarray}
{\cal S}^{2}-2\pi\left(2M^2-Q^2 \right){\cal S}+ \left(2\pi J\right)^2 &=& 0 ~.\label{quad2}
\end{eqnarray}
It indicates that ``entropy product'' is independent of mass and `entropy sum''
depends on the BH mass.

Using Eq. (\ref{aread}), one can derive another important relations:
\begin{eqnarray}
T_{+}{\cal S}_{+}+ T_{-}{\cal S}_{-} &=& 0 ~.\label{exs1}
\end{eqnarray}
and
\begin{eqnarray}
\frac{{\Omega}_{+}}{T_{+}}+\frac{{\Omega}_{-}}{T_{-}} &=& 0 ~. \label{oms}
\end{eqnarray}
The above important thermodynamic products of multi horizons may be used to determine
the classical BH entropy in terms of Cardy formula, therefore giving some evidence for a 
BH/CFT description of the corresponding microstates\cite{castro12}. It has been also shown that
from the above Eq.(\ref{exs1}), the central charge being the same for two horizon BHs. Explicit 
calculation of the central charges $c_{L}=c_{R}=12J$ using Cardy formula have been done in 
Appendix B. Using thermodynamical relations, we derive the dimensionless temperature of 
microscopic CFT, which is perfect agreement with the ones derived from hidden conformal symmetry 
in the low frequency scattering off the BH\cite{sia13}.

Based on these above relations, we would like to compute the entropy bound of ${\cal H}^{\pm}$ 
which is exactly Penrose-like in-equality  for event horizon. 
From the Eq. \ref{ieqs}, we obtain  Kerr like bound for Sen BH:
\begin{eqnarray}
M^4-Q^2M^2 + \frac{Q^4-4J^2}{4} \geq 0 ~.\label{ineq1}
\end{eqnarray}
or
\begin{eqnarray}
M^2 \geq J+ \frac{Q^2}{2} ~.\label{ieq}
\end{eqnarray}
Since $r_{+} \geq r_{-}$  thus ${\cal S}_{+} \geq {\cal S}_{-} \geq 0$.
Then the entropy product (\ref{etpsen}) gives:
\begin{eqnarray}
{\cal S}_{+} \geq  \sqrt{{\cal S}_{+} {\cal S}_{-}}= 2\pi J \geq {\cal S}_{-} ~.\label{ieq1}
\end{eqnarray}
and the entropy sum gives:
\begin{eqnarray}
2\pi \left(2 M^2-Q^2\right) &=& {\cal S}_{+}+ {\cal S}_{-} \geq {\cal S}_{+}
\geq \frac{{\cal S}_{+}+ {\cal S}_{-}}{2}= \pi \left(2M^2-Q^2\right) \geq {\cal S}_{-}  ~.\label{inq2}
\end{eqnarray}
Thus the entropy bound for  ${\cal H}^{+}$:
\begin{eqnarray}
 \pi \left(2M^2-Q^2\right)  \leq {\cal S}_{+} \leq 2\pi \left(2M^2-Q^2\right)   ~.\label{inq3}
\end{eqnarray}
and  the entropy bound for  ${\cal H}^{-}$:
\begin{eqnarray}
 0 \leq {\cal S}_{-} \leq   2\pi J ~.\label{inq4}
\end{eqnarray}
From this bound,  we can derive area bound which could be found in the latter section. It should be noted that 
in the limit $Q=0$, we obtain the Kerr entropy bound\cite{xu}.

Similarly, we can obtain the ``product of surface gravity'' and ``sum of surface
gravity'' of ${\cal H}^{\pm}$  is
\begin{eqnarray}
{\kappa}_{-} {\kappa}_{+} &=& -\frac{(2M^2-Q^2)^{2}-4J^{2}}{(4JM)^{2}}~.\label{psg}
\end{eqnarray}
and
\begin{eqnarray}
{\kappa}_{-} +{\kappa}_{+} &=& -\frac{(2M^2-Q^2)^{2}-4J^{2}}{4MJ^{2}}~.\label{psg1}
\end{eqnarray}
It suggests that  surface gravity product and  surface gravity sum are not
universal.
It may be noted that surface gravity satisfied the following quadratic equation.
\begin{eqnarray}
{\kappa}^{2}-\left(\frac{4J^{2}-(2M^2-Q^2)^{2}}{4MJ^{2}}\right){\kappa}+
\left(\frac{4J^{2}-(2M^2-Q^2)^{2}}{4MJ^{2}}\right) &=& 0 ~.\label{kapq}
\end{eqnarray}

Similarly,  one can obtain ``surface temperature product'' and
``surface temperature sum'' of ${\cal H}^{\pm}$ as follows
\begin{eqnarray}
{T}_{-}{T}_{+} &=&-\frac{(2M^2-Q^2)^{2}-4J^{2}}{(8\pi JM)^{2}} ~.\label{pst1}
\end{eqnarray}
and
\begin{eqnarray}
{T}_{-}+{T}_{+} &=&-\frac{(2M^2-Q^2)^{2}-4J^{2}}{8\pi M J^{2}} ~.\label{pst2}
\end{eqnarray}
It seems that these products and sum are \emph{not} universal.

Finally, ``Komar energy product'' and ``Komar energy sum''  of ${\cal H}^{\pm}$
for Sen BH is given by
\begin{eqnarray}
E_{+}E_{-}  &=& (2{\cal S}_{+} T_{+}) (2 {\cal S}_{-} T_{-})
=-\left[(2M^2-Q^2)^{2}-4J^{2} \right] ~. \label{kengy1}
\end{eqnarray}
and
\begin{eqnarray}
E_{+}+E_{-}  &=& (2{\cal S}_{+} T_{+})+ (2 {\cal S}_{-} T_{-})
=0 ~. \label{kengy2}
\end{eqnarray}

The above calculation suggests that the product of the area and entropy of ${\cal H}^{\pm}$   are proportional
to the square of the spin parameter $J$. Surface gravity product, surface temperature product and Komar energy
product depends on  ADM mass. Thus, we may conclude that they are not universal except the area product
and entropy product.
In appendix A, we have computed various thermodynamic parameters for Kerr-Newman BH, Kerr BH
in comparison with Sen BH. Now we are going to derive the Smarr formula for Sen BH.

\subsection{\label{smar} Smarr Formula for Sen BH:}
It is well known that for KN BH  the area of the outer\cite{smarr73} and inner horizons are
\begin{eqnarray}
{\cal A}_{\pm} &=& 4\pi \left( 2M^2-Q^2 \pm 2\sqrt{M^4-J^2-M^2Q^2}\right) ~.\label{arKN}
\end{eqnarray}

Indeed, it is  constant over the ${\cal H}^{\pm}$. Similarly, we can evaluate the area
of ${\cal H}^{\pm}$ for Sen BH reads
\begin{eqnarray}
{\cal A}_{\pm} &=& 8\pi M \left[\left(M-\frac{Q^2}{2M}\right) \pm \sqrt{\left(M-\frac{Q^2}{2M}\right)^{2}-a^2}\right]
~.\label{areasen1}
\end{eqnarray}

Inverting the above relation one can compute the BH mass or ADM mass can be expressed
in terms of area of both the horizon.
\begin{eqnarray}
M^2 &=& \frac{{\cal A}_{\pm}}{16\pi}+\frac{4\pi J^2}{{\cal A}_{\pm}}
+\frac{Q^2}{2} ~.\label{mas}
\end{eqnarray}

It is remarkable  that the mass can be expressed as in terms of both area of
${\cal H}^{+}$ and ${\cal H}^{-}$. Now we will see what happens with the mass
differential?.  It could be also expressed as three physical invariants of
both ${\cal H}^+$ and ${\cal H}^-$,
\begin{eqnarray}
dM &=& \Gamma_{\pm} d{\cal A}_{\pm} + \Omega_{\pm} dJ +\Phi_{\pm} dQ
~. \label{dm}
\end{eqnarray}
where
\begin{eqnarray}
\Gamma_{\pm} &=& \frac{\partial M}{\partial {\cal A}_{\pm}}= \frac{1}{M} 
\left(\frac{1}{32 \pi}-\frac{2\pi J^2}{{\cal A}_{\pm}^2}\right)\nonumber \\
\Omega_{\pm} &=& \frac{\partial M}{\partial J}=\frac{4\pi J}{M{\cal A}_{\pm}}
=\frac{a}{2Mr_{\pm}} \nonumber\\
\Phi_{\pm} &=& \frac{\partial M}{\partial Q}=\frac{Q}{M}   ~. \label{invar}
\end{eqnarray}
where
\begin{eqnarray}
\Gamma_{\pm} &=& \mbox{Effective surface tension for ${\cal H}^{+}$ and ${\cal H}^{-}$}
\nonumber \\
\Omega_{\pm} &=&  \mbox{Angular velocity for ${\cal H}^\pm$} \nonumber \\
\Phi_{\pm} &=& \mbox{Electromagnetic potentials for ${\cal H}^\pm$}\nonumber
\end{eqnarray}

The effective surface tension can be rewritten as
\begin{eqnarray}
\Gamma_{\pm} &=& \frac{1}{ M} \left(\frac{1}{32 \pi}-\frac{2\pi J^2}{{\cal A}_{\pm}^2} \right)\\
&=& \frac{1}{32 \pi M} \left(1-\frac{64\pi^2J^2}{{\cal A}_{\pm}^2}\right)\nonumber\\
&=& \frac{1}{32 \pi M} \left(1-\frac{a^2}{r_{\pm}^2}\right)\nonumber\\
&=& \frac{r_{\pm}-M}{32\pi Mr_{\pm}}= \frac{{\kappa}_{\pm}} {8\pi}
\end{eqnarray}
where $\kappa_{\pm}$ is the surface gravity of ${\cal H}^{\pm}$ as previously defined.

Thus the mass can be expressed in terms of these quantities both for
${\cal H}^\pm$ as a simple bilinear form
\begin{eqnarray}
M &=& 2\Gamma_{\pm} {\cal A}_{\pm} + 2J \Omega_{\pm} +2Q \Phi_{\pm}
~. \label{bilinear}
\end{eqnarray}

This has been derived from the homogenous function of degree $\frac{1}{2}$ in
$({\cal A}_{\pm}, J, Q)$.
Remarkably, $\Gamma_{\pm}$,  $\Omega_{\pm}$ and $\Phi_{\pm}$  are constant
on the ${\cal H}^+$ and ${\cal H}^-$ for any stationary, axially symmetric
space-time.

Since the $dM$ is a total perfect differential, one may choose freely any path of
integration in $({\cal A}_{\pm}, J, Q)$ space. Thus one could define
surface energy ${\cal E}_{s, \pm}$ for ${\cal H}^{\pm}$
\begin{eqnarray}
{\cal E}_{s, \pm} &=& \int_{0}^{{\cal A}_{\pm}} \Gamma (\tilde{{\cal A}_{\pm}}
, 0 ,0) d\tilde{{\cal A}_{\pm}}; ~ \label{se}
\end{eqnarray}
the rotational energy  for ${\cal H}^{\pm}$  can be defined by
\begin{eqnarray}
{\cal E}_{r, \pm} &=& \int_{0}^{J} \Omega_{\pm} ({\cal A}_{\pm}
, \tilde{J} ,0) d\tilde{J},\,\,  \mbox{${\cal A}_{\pm}$ fixed}; ~ \label{re}
\end{eqnarray}

and the electromagnetic energy  for ${\cal H}^{\pm}$ is
\begin{eqnarray}
{\cal E}_{em, \pm} &=& \int_{0}^{Q} \Phi ({\cal A}_{\pm}
, J, \tilde{Q}) d\tilde{Q},\,\, \mbox{${\cal A}_{\pm}$, $J$  fixed}; ~ \label{re1}
\end{eqnarray}

Therefore, we may rewrite the Eq. (\ref{bilinear}) as
\begin{eqnarray}
M &=& \pm \frac{{\kappa}_{\pm}}{4\pi}{\cal A}_{\pm} + 2J\Omega_{\pm}+2Q\Phi_{\pm}
~. \label{bilinear1}
\end{eqnarray}
or
\begin{eqnarray}
 M-2J\Omega_{\pm}-2Q\Phi_{\pm}  &=& \pm \frac{{\kappa}_{\pm}}{4\pi}{\cal A}_{\pm}
~. \label{bilinear2}
\end{eqnarray}
or
\begin{eqnarray}
 M-2J\Omega_{\pm}-2\Phi_{\pm} Q &=&\pm \frac{T_{\pm}}{2}{\cal A}_{\pm}
~. \label{bilinear3}
\end{eqnarray}
or
\begin{eqnarray}
\frac{M}{2} &=& \pm {T}_{\pm}{\cal S}_{\pm} + J\Omega_{\pm}+Q \Phi_{\pm}
~. \label{bilinear4}
\end{eqnarray}
This could be  recognized as a generalized \emph{Smarr-Gibbs-Duhem} relation on
${\cal H}^{\pm}$ for Sen BH.


\subsection{\label{ruffini} Irreducible Mass Product for Sen BH:}

In this section, we will derive Christodoulou-Ruffini\cite{cr71} mass formula for
Sen BH. Christodoulou had shown that the irreducible mass $M_{irr}$ of a Kerr
BH is related to the surface area ${\cal A}$ of the BH by the following formula
\begin{eqnarray}
M_{irr}^{2} &=& \frac{{\cal A}}{16\pi}
~. \label{irrm}
\end{eqnarray}
It is now well known that this formula is valid for both the horizons.  Thus we can define it for
${\cal H}^{\pm}$:
\begin{eqnarray}
M_{irr, \pm}^{2} &=& \frac{{\cal A}_{\pm}}{16\pi}=\frac{Mr_{\pm}}{2}
~. \label{irrm1}
\end{eqnarray}
where `$+$' sign indicates for ${\cal H}^{+}$ and `$-$' indicates for ${\cal H}^{-}$.

Likewise, the area and angular velocity may be expressed in terms of
$M_{irr, \pm}$:

\begin{eqnarray}
 {\cal A}_{\pm} &=& 16 \pi (M_{irr, \pm})^2
~. \label{irrma}
\end{eqnarray}

and
\begin{eqnarray}
 {\Omega}_{\pm} &=&  \frac{a}{4(M_{irr,\pm})^2}
~. \label{iromega}
\end{eqnarray}
Interestingly, the product of the irreducible mass of ${\cal H}^{\pm}$ for
Sen BH is \emph {universal}.
\begin{eqnarray}
M_{irr,+} M_{irr,-} &=& \frac{J}{2} ~. \label{irrmp1}
\end{eqnarray}
The Christodoulou-Ruffini mass formula for Sen BH and for both the horizon
(${\cal H}^{\pm}$) reads as:
\begin{eqnarray}
M^2 &=& \left(M_{irr,\pm}+\frac{Q^2}{4M_{irr,\pm}}\right)+\frac{J^2}{4 (M_{irr,\pm})^2} ~. \label{irrma1}
\end{eqnarray}

Based on the above relations, we would like to compute the area bound and irreducible mass bound for Sen
BH followed by the previous sections. 
Since $r_{+} \geq r_{-}$, one obtains ${\cal A}_{+} \geq {\cal A}_{-} \geq 0$.
Therefore the area product gives:
\begin{eqnarray}
{\cal A}_{+}  \geq  \sqrt{{\cal A}_{+} {\cal A}_{-}}= 8 \pi J
\geq {\cal A}_{-}
~.\label{ieqa1}
\end{eqnarray}
and the area sum gives:
\begin{eqnarray}
8\pi \left(2 M^2-Q^2\right) = {\cal A}_{+}+ {\cal A}_{-} \geq {\cal A}_{+}
\geq \frac{{\cal A}_{+}+ {\cal A}_{-}}{2}= 4\pi \left(2M^2-Q^2 \right) \geq {\cal A}_{-} ~.\label{inqa2}
\end{eqnarray}
Thus the area bound for  ${\cal H}^{+}$:
\begin{eqnarray}
 4\pi \left(2M^2-Q^2\right)  \leq {\cal A}_{+} \leq 8\pi \left(2M^2-Q^2\right)   ~.\label{inqa3}
\end{eqnarray}
and  the area bound for  ${\cal H}^{-}$:
\begin{eqnarray}
 0 \leq {\cal A}_{-} \leq  8 \pi J ~.\label{inqa4}
\end{eqnarray}
From this area bound, we get irreducible mass bound for Sen BH:
For ${\cal H}^{+}$:
\begin{eqnarray}
\frac{\sqrt{2M^2-Q^2}}{2} \leq M_{irr, +} \leq \frac{\sqrt{2M^2-Q^2}}{\sqrt{2}}   ~.\label{inqa5}
\end{eqnarray}
and  for ${\cal H}^{-}$:
\begin{eqnarray}
0 \leq M_{irr,-} \leq  \sqrt{\frac{J}{2}}~.\label{inqa6}
\end{eqnarray}
Eq. \ref{inqa5} is nothing but the Penrose inequality, which is the first geometric inequality for 
BHs\cite{peni}.

\subsection{\label{law} The four Laws of BH Thermodynamics on ${\cal H}^{\pm}$ :}

Let us quickly examine the four laws of BH thermodynamics for Sen BH. For Kerr-Newman BH,
 Carter, Hawking and Bardeen \cite{bcw73} formulated the black hole thermodynamics  for the event horizon
which is analogous to the classical laws of thermodynamics. We  derive here same for Sen BH both
on the EH as well as CH. We have already been derived the surface gravity in the previous
section given by the Eq. (\ref{sgKN}) and Eq. (\ref{tmKN}). Using two Eqs. we can easily say that 
the surface gravity and the surface temperature  are constant on the ${\cal H}^{\pm}$ and 
therefore, it is remarkable that the Zeroth law of BH thermodynamics holds for CH as well as EH.

\begin{itemize}
\item The Zeroth Law: The surface gravity, $\kappa_{\pm}$ of a stationary black hole is constant over  the
event horizon (${\cal H }^{+}$) as well as Cauchy horizon (${\cal H }^{-}$).

Quite similarly, the first law of BH thermodynamics is also satisfied not only at the  outer horizon  but
also at the inner horizon, that is

\item The First Law: Any perturbation of a stationary BHs, the change of mass (change of energy) is
related to change of mass, angular momentum, and electric charge by:
\begin{eqnarray}
dM &=& \pm \frac{{\kappa}_{\pm}} {8\pi} d{\cal A}_{\pm} + \Omega_{\pm} dJ +\Phi_{\pm}dQ ~. \label{dms2}
\end{eqnarray}
It can be seen that $\frac{{\kappa}_{\pm}} {8\pi}$ is analogous to the temperature of ${\cal H}^{\pm}$  
in the same way that ${\cal A}_{\pm}$ is analogous to entropy. It should be noted that 
$\frac{{\kappa}_{\pm}} {8\pi}$ and ${\cal A}_{\pm}$
are quite distinct from the temperature and entropy of the BH.

Again, the second law of BH thermodynamics is also satisfied both on the inner horizon and outer horizon,
that is:

\item The Second Law: The area  ${\cal A}_{\pm}$ of both event horizons $({\cal H}^{+})$ and
 Cauchy horizons $({\cal H}^{-})$ never decreases, i.e.
 \begin{eqnarray}
 d{\cal A}_{\pm} &=& \frac{4 {\cal A}_{\pm}}{r_{\pm}-r_{\mp}} \left(dM-\vec{\Omega}_{\pm}.d\vec{J}
-\Phi_{\pm} dQ \right)\geq 0 \label{arth}
\end{eqnarray}
or
\begin{eqnarray}
 dM_{irr, \pm}&=& \frac{2M_{irr, \pm}}{r_{\pm}-r_{\mp}} \left(dM-\vec{\Omega}_{\pm}.d\vec{J}
-\Phi_{\pm} dQ \right) \geq 0 \label{arth1}
\end{eqnarray}
The change in irreducible mass of both event horizons $({\cal H}^{+})$ and Cauchy horizons $({\cal H}^{-})$ can
never be negative. It follows that immediately
\begin{eqnarray}
  dM > \vec{\Omega}_{\pm}.d\vec{J}+\Phi_{\pm} dQ  \label{srad}
\end{eqnarray}

For the extremal Sen BH($r_{+}=r_{-}$), we have $T_{+}=T_{-}=0=\kappa_{+}=\kappa_{-}$. Therefore the third law
becomes:
\item The Third Law: It is impossible by any mechanism, no matter how idealized, to reduce, $\kappa_{\pm}$ the surface
gravity of  both event horizon $({\cal H}^{+})$ and Cauchy horizon $({\cal H}^{-})$ to zero by a finite number of
operations.
\end{itemize}

Thus we have checked that the four laws of BH mechanics satisfied on CH as well as EH.

So far all the computations  have been carried out for Sen BH in ``Einstein frame (EF)''. Now we
will see in next section, what happens for these computations for Sen BH in ``String frame (SF)''?

\section{\label{sen} Sen BH in String Frame: }

This frame sometimes used because in this frame the physical degrees of  freedom
move along the geodesics of the metric\cite{harm99}. Therefore the corresponding
metric in the SF and the EF are conformally  related by the following relation
\begin{eqnarray}
G_{ab} &=& e^{2\phi} g_{ab} ~. \label{gab}
\end{eqnarray}
where $G_{ab}$ are the covariant components of the metric in the SF, $g_{ab}$
are the components of the metric in the EF and $\phi$ is the dilation field.
For contravariant components they are related by
\begin{eqnarray}
G^{ab} &=& e^{-2\phi} g^{ab} ~. \label{gab1}
\end{eqnarray}

The dilation field is given by
\begin{eqnarray}
e^{2\phi}  &=& \frac{r^2+a^2 \cos^2\theta}{\rho^{2}} ~. \label{cf}
\end{eqnarray}
For simplicity, we denote
\begin{eqnarray}
\chi  &=& r^2+a^2 \cos^2\theta ~. \label{cf1}
\end{eqnarray}

First, we need to write the metric for Sen BH in String frame \cite{blaga01}, which is given by
$$
ds^2 = -\frac{\chi}{\rho^{2}}\left(1-\frac{2mr \cosh^2\alpha }{\rho^2}\right)dt^2-
\frac{4amr\chi \cosh^2 \alpha \sin^2\theta}{\rho^4}dt d\phi +\frac{\chi}{\Delta}dr^2+\chi d\theta^2
$$
\begin{eqnarray}
+ \frac{\chi}{\rho^4} \Upsilon \sin^2\theta d\phi^2 ~\label{sen1}
\end{eqnarray}
Since the action in SF is different from EF therefore the conserved quantities are also different. Now we define 
this conserved quantities(mass, charge, angular momentum) in SF are ${\cal M}$, ${\cal Q}$ and ${\cal J}$ respectively. 
Then the horizon radii in SF becomes 
\begin{eqnarray}
r_{\pm}^{SF}&=& \left({\cal M}-\frac{{\cal Q}^2}{2{\cal M}}\right)\,\, \pm 
\sqrt{\left({\cal M}-\frac{{\cal Q}^2}{2{\cal M}}\right)^{2}-a^2}
~.\label{eq1}
\end{eqnarray}
Here the spin parameter $a=\frac{{\cal J}}{{\cal M}}$.

Now the area of both the horizons (${\cal H}^\pm$) in SF is given by
\begin{eqnarray}
{{\cal A}_{\pm}}^{SF} &=& 4\pi \left[r_{\pm}^{SF}(r_{\pm}^{SF}+b)+a^2\right]
               \left[1-\frac{b r_{\pm}^{SF}}{a\sqrt{r_{\pm}^{SF}(r_{\pm}^{SF}+b)}}
               \tan^{-1}\frac{a}{\sqrt{r_{\pm}^{SF}(r_{\pm}^{SF}+b)}}  \right]
~.\label{areaSF}
\end{eqnarray}
where, $b=\frac{{\cal Q}^2}{{\cal M}}$, $G_{\theta\theta}=\chi$ and
$G_{\phi\phi}=\frac{\chi}{\rho^{4}} \Upsilon \sin^2\theta$.
Similarly, we can compute the entropy for both the horizons (${\cal H}^{\pm}$) in the SF:
\begin{eqnarray}
{{\cal S}_{\pm}}^{SF} &=& \pi \left[r_{\pm}^{SF}(r_{\pm}^{SF}+b)+a^2\right] 
\left[1-\frac{r_{\pm}^{SF}b}{a\sqrt{r_{\pm}^{SF}(r_{\pm}^{SF}+b)}}
\tan^{-1}\frac{a}{\sqrt{r_{\pm}^{SF}(r_{\pm}^{SF}+b)}}  \right] ~.\label{entropySF}
\end{eqnarray}
Now we turn to the most interesting case that is the ``Area product'' for Sen BH
in SF:
$$
{{\cal A}_{+}}^{SF}{{\cal A}_{-}}^{SF} =  \left(8\pi {\cal J}\right)^2 
\left[1-\frac{b r_{+}^{SF}}{a\sqrt{r_{+}^{SF}(r_{+}^{SF}+b)}}\tan^{-1}\frac{a}
{\sqrt{r_{+}^{SF}(r_{+}^{SF}+b)}}\right] \times
$$
\begin{eqnarray}
\left[1-\frac{b r_{-}^{SF}}{a\sqrt{r_{-}^{SF}(r_{-}^{SF}+b)}}\tan^{-1}\frac{a}
{\sqrt{r_{-}^{SF}(r_{-}^{SF}+b)}}\right] ~.\label{arSF}
\end{eqnarray}
Interestingly, it seems that the product of horizon area of ${\cal H}^{\pm}$ in SF for Sen BH
is  \emph{not universal}. This is one of the key results of the work. The results in the SF is quite 
different from the EF due to the fact that the action in SF is quite different from the EF, therefore 
the corresponding conserved quantities should be different. Actually,  when the $M$, $J$ 
and $Q$ are computed in EF the action should be Einstein-Hilbert type and where the ADM formulas have 
been used, whereas when the action is written in SF, the corresponding quantities are very likely to be 
different. Therefore the parameters $M$, $J$ in particular could no longer be identified with 
the conserved charges associated with the time-translation and the rotational symmetry. 

But if we expand the function of $tan^{-1}x$ as
\begin{eqnarray}
 tan^{-1} x &=& x-\frac{x^3}{3}+\frac{x^5}{5}-\frac{x^7}{7}+...~.\label{eq2}
\end{eqnarray}
then we find the the area of both the horizons (${\cal H}^\pm$) as
\begin{eqnarray}
{{\cal A}_{\pm}}^{SF} &=& \frac{8\pi{\cal M} (r_{\pm}^{SF})^2}{(r_{\pm}^{SF}+b)}
\left[1+\frac{b}{3(r_{\pm}^{SF}+b)}\left(\frac{a}{r_{\pm}^{SF}}\right)^2-
 \frac{b}{5(r_{\pm}^{SF}+b)^2}\left(\frac{a}{r_{\pm}^{SF}}\right)^4 
 +{\cal O}\left(\frac{a}{r_{\pm}^{SF}}\right)^6\right] ~.\label{araSF}\nonumber
\end{eqnarray}
It follows from the above equation it is very difficult to find the exact mass parameter in terms of the
area of ${\cal H}^\pm$ in SF. Therefore due to same reasons it is also quite difficult to find the Hawking 
temperature from the mass differential. So one way we could find the Hawking temperature in SF by using the 
formula as used Sen in\cite{as92}:
\begin{eqnarray}
T &=& \frac{\kappa}{2\pi}
=\frac{\lim_{r\rightarrow r_{\pm}^{SF}}\sqrt{G^{rr}}\partial_{r}\sqrt{-G_{tt}}}{2\pi}|_{\theta=0}
~.\label{eq3}
\end{eqnarray}
which gives on the ${\cal H}^\pm$
\begin{eqnarray}
T_{\pm} &=& \frac{2(r_{\pm}^{SF}-{\cal M})+b}{4\pi\left[r_{\pm}^{SF}(r_{\pm}^{SF}+b)+a^2\right]}
~.\label{eq4}
\end{eqnarray}

Similarly, we could find the angular velocity by using the formula
\begin{eqnarray}
 \Omega^{SF} &=& \frac{-G_{t\phi}+\sqrt{G_{t\phi}^2-G_{\phi\phi}G_{tt}}}{G_{\phi\phi}} ~.\label{eq5}
\end{eqnarray}
On the horizon the angular velocity could be written as 
\begin{eqnarray}
 \Omega_{\pm}^{SF} &=&- \frac{G_{t\phi}}{G_{\phi\phi}}=\frac{2a{\cal M}r_{\pm}^{SF}}
 {\left[r_{\pm}^{SF}(r_{\pm}^{SF}+b)+a^2\right]^2} ~.\label{eq6}
\end{eqnarray}
Now we can write the first law of thermodynamics in the SF as 
\begin{eqnarray}
d{\cal M} &=&\pm T_{\pm}^{SF} d{\cal S}_{\pm}^{SF} + \Omega_{\pm}^{SF} d{\cal J}+... ~\label{eq7}
\end{eqnarray}
Now it implies that the BH temperature, angular velocity and probably electric potentials(since charge is different) 
in SF are quite different from EF because the action and metric are different as we have discussed previously. 
This is why the area(or entropy) product relation in two frames are quite \emph{distinct}.

We also note that the sum of horizon area in SF  reads
$$
{{\cal A}_{+}}^{SF}+{{\cal A}_{-}}^{SF} = {8\pi{\cal M} r_{+}}^{SF}
\left[1-\frac{br_{+}^{SF}}{a\sqrt{r_{+}^{SF}(r_{+}^{SF}+b)}} \tan^{-1}\frac{a}
{\sqrt{r_{+}^{SF}(r_{+}^{SF}+b)}}\right]+
$$
\begin{eqnarray}
{8\pi{\cal M} r_{-}}^{SF}
\left[1-\frac{b r_{-}^{SF}}{a\sqrt{r_{-}^{SF}(r_{-}^{SF}+b)}} \tan^{-1}\frac{a}
{\sqrt{r_{-}^{SF}(r_{-}^{SF}+b)}}\right]
~.\label{arsumSF}
\end{eqnarray}

Like-wise, the entropy product and entropy sum for Sen BH in SF is
$$
{{\cal S}_{+}}^{SF}{{\cal S}_{-}}^{SF} = \left(2\pi {\cal J}\right)^2
 \left[1-\frac{b r_{+}^{SF}}{a\sqrt{r_{+}^{SF}(r_{+}^{SF}+b)}} \tan^{-1}\frac{a}
{\sqrt{r_{+}^{SF}(r_{+}^{SF}+b)}}\right] \times
$$
\begin{eqnarray}
\left[1-\frac{b r_{-}^{SF}}{a\sqrt{r_{-}^{SF}(r_{-}^{SF}+b)}} \tan^{-1}\frac{a}
{\sqrt{r_{-}^{SF}(r_{-}^{SF}+b)}}\right]
\end{eqnarray}
and
$$
{{\cal S}_{+}}^{SF}+{{\cal S}_{-}}^{SF} ={2\pi{\cal M} r_{+}}^{SF}
 \left[1-\frac{b r_{+}^{SF}}{a\sqrt{r_{+}^{SF}(r_{+}^{SF}+b)}} 
\tan^{-1}\frac{a}{\sqrt{r_{+}^{SF}(r_{+}^{SF}+b)}}\right]+
$$
\begin{eqnarray}
{2\pi{\cal M} r_{-}}^{SF}
\left[1-\frac{b r_{-}^{SF}}{a\sqrt{r_{-}^{SF}(r_{-}^{SF}+b)}} \tan^{-1}\frac{a}
{\sqrt{r_{-}^{SF}(r_{-}^{SF}+b)}}\right]
\end{eqnarray}
It also implies that the  entropy product and entropy sum for ${\cal H}^{\pm}$ in SF
of Sen BH are  \emph{not universal}.

For completeness, we also compute the  irreducible mass of Sen BH for ${\cal H}^{\pm}$
in SF reads
\begin{eqnarray}
 {{\cal M}_{irr,\pm}}^{SF}
&=& \sqrt{\frac{{\cal M}r_{\pm}^{SF}}{2}\left[1-\frac{b r_{\pm}^{SF}}{a\sqrt{r_{\pm}^{SF}(r_{\pm}^{SF}+b)}} 
\tan^{-1}\frac{a}{\sqrt{r_{\pm}^{SF}(r_{\pm}^{SF}+b)}}  \right]}  ~. \label{irrsf}
\end{eqnarray}
For our record, we find the irreducible mass product in SF:
$$
{{\cal M}_{irr,+}}^{SF} {{\cal M}_{irr,-}}^{SF} =\frac{{\cal J}}{2} 
\sqrt{\left[1-\frac{b r_{+}^{SF}}{a\sqrt{r_{+}^{SF}(r_{+}^{SF}+b)}}\tan^{-1}\frac{a}{\sqrt{r_{+}^{SF}(r_{+}^{SF}+b)}}\right]} 
\times
$$
\begin{eqnarray}
\sqrt{\left[1-\frac{r_{-}^{SF}b}{a\sqrt{r_{-}^{SF}(r_{-}^{SF}+b)}} 
\tan^{-1}\frac{a}{\sqrt{r_{-}^{SF}(r_{-}^{SF}+b)}}\right]} ~.\label{irrsf2}
\end{eqnarray}
and the sum of irreducible mass is
$$
{{\cal M}_{irr,+}}^{SF}+ {{\cal M}_{irr,-}}^{SF}=\sqrt{\frac{{\cal M}r_{+}^{SF}}{2}}
\sqrt{\left[1-\frac{b r_{+}^{SF}}{a\sqrt{r_{+}^{SF}(r_{+}^{SF}+b)}}
\tan^{-1}\frac{a}{\sqrt{r_{+}^{SF}(r_{+}^{SF}+b)}}\right]}+
$$
\begin{eqnarray}
\sqrt{\frac{{\cal M}r_{-}^{SF}}{2}} \sqrt{\left[1-\frac{b r_{-}^{SF}}{a\sqrt{r_{-}^{SF}(r_{-}^{SF}+b)}} 
\tan^{-1}\frac{a}{\sqrt{r_{-}^{SF}(r_{-}^{SF}+b)}} \right]} ~. \label{irrs4}
\end{eqnarray}
It seems  that they both are \emph{not universal}. It is obvious because irreducible mass depends on area. 
It is true that area product, entropy product and irreducible mass product gives same result because
they are identical. For our record we computed them separately.
\section{\label{dis} Discussion}
In this work, we have examined various thermodynamic products for rotating charged black hole
solution in four dimensional heterotic string theory. We have considered both  EF and SF.
In the EF, we have shown that  the ``area product'' and ``entropy product'' are universal,
while the ``area sum'' and the``entropy sum'' are not! In the SF, we  have shown that the ``area product'',
``entropy product'',``area sum'' and ``entropy sum'' do not manifested any universal character because
they  all  are depends on ADM mass parameter. We also showed that every BH thermodynamical
variable, other than the mass ($M$), the angular momentum ($J$) and the charge $(Q)$ parameter,
 can form a quadratic equation whose roots are contained the three basic parameter $M,J,Q$.
For completeness, we have derived the Smarr mass formula and Christodoulou's irreducible mass 
formula for Sen BH in the EF. Finally, we showed that the four laws of BH mechanics satisfied 
on both the horizons ${\cal H}^{\pm}$.

Based on the thermodynamic relations, we also derived the  area bound and entropy bound
for all the horizons. Furthermore, we calculated the irreducible mass bound for this type of BH. 
These formulas are expected to be useful to understanding  the microscopic nature of BH entropy (both 
exterior and interior). Again, the entropy products of inner horizon and outer horizons could be 
used to determine whether the classical BH entropy could be written as a Cardy formula(See Appendix B), 
giving some evidence for a holographic description of BH/CFT correspondence\cite{chen12}. The above thermodynamic 
properties including the Hawking temperature and area of both the horizons may therefore be expected 
to play a crucial role to understanding the BH entropy at the microscopic level.

There has been compelling evidence by astrophysically that BH's have EH\cite{ramesh} and it is 
also true that the EH's are thermodynamically stable with respect to axi-symmetric perturbations
\cite{kaburki91,kaburki92}. Whereas there is no strong or weak evidence that BH's have CH by 
astrophysically but analytically has strong evidence that BH's possesses CH in addition with EH, 
and it is also well known by fact that CH is thermodynamically unstable by axi-symmetric 
perturbations\cite{frolov89}. So still now it is unclear to us whether the CH thermodynamic 
results have a real astrophysical significance just as the event horizon does\cite{ramesh}. 
So, it will be a little bit help us to understanding the interior physics of Sen BH to clarify what 
the inner thermodynamics physically means.  It may quite plausible that this interior physics 
could help us to understanding the interior BH entropy.

\appendix
\section{Appendix:}
\begin{center}
\begin{tabular}{|c|c|c|c|}
    \hline
    Parameter & KN BH & Kerr BH  & Sen BH\\
    \hline
    $r_{\pm}$: &$M\pm\sqrt{M^{2}-a^2-Q^{2}}$ &$M\pm\sqrt{M^{2}-a^2}$ & $ M-\frac{Q^2}{2M}\pm \sqrt{(M-\frac{Q^2}{2M})^{2}-a^2}$\\

    $\sum r_{i}$: & $2M$ & $2M$ & $2M-\frac{Q^2}{M}$\\

    $\prod r_{i}$: & $a^2+Q^2$ & $a^2$ & $a^2$ \\

    ${\cal A}_{\pm}$: & $4\pi \left(r_{\pm}^2+a^2\right)$ & $4\pi \left(2M^2 \pm 2\sqrt{M^4-J^2}\right)$ & $8\pi Mr_{\pm}$ \\

    $\sum {\cal A}_{i}$ : &$8\pi\left(2M^2-Q^2\right)$ &$16\pi M^2$& $8\pi\left(2M^2-Q^2\right)$\\

    $\prod {\cal A}_{i}$: &$(8\pi)^2\left(J^2+\frac{Q^4}{4}\right)$ & $\left(8\pi J\right)^2$ & $\left(8\pi J\right)^2 $ \\

    $S_{\pm}$: &$\pi(r_{\pm}^2+a^2)$ & $\pi(r_{\pm}^2+a^2)$ & $2\pi Mr_{\pm} $\\

    $\sum {\cal S}_{i}$ : &$2\pi(2M^2-Q^2)$& $4\pi M^2$ & $2\pi\left(2M^2-Q^2 \right)$\\

    $\prod {\cal S}_{i}$: & $(2\pi)^2\left(J^2+\frac{Q^4}{4}\right)$&$\left(2\pi J\right)^2$&
    $\left(2\pi J\right)^2$\\

    $\kappa_{\pm}$: &$ \frac{r_{\pm}-r_{\mp}}{2(r_{\pm}^2+a^2)}$&$ \frac{r_{\pm}-r_{\mp}}{2(r_{\pm}^2+a^2)}$&$\frac{r_{\pm}-r_{\mp}}{4Mr_{\pm}}$\\

    $\sum{\kappa}_{i}$: &$\frac{4M(a^2+Q^2-M^2)}{(4J^2+Q^4)} $&$\frac{a^2-M^2}{aJ}$&$\frac{4J^{2}-(2M^2-Q^2)^{2}}{4MJ^{2}} $\\

    $\prod {\kappa}_{i}$: &$\frac{a^2+Q^2-M^2}{4M^2(a^2+Q^2)} $&$\frac{a^2-M^2}{4J^2}$&$\frac{4J^{2}-(2M^2-Q^2)^{2}}{(4MJ)^{2}}$\\

    $T_{\pm}$:   &$\frac{r_{\pm}-r_{\mp}}{4\pi (r_{\pm}^2+a^2)}$&$\frac{r_{\pm}-r_{\mp}}{4\pi (r_{\pm}^2+a^2)}$&$ \pm \frac{\sqrt{(2M^2-Q^2)^2-4J^2}}{4\pi M[(2M^2-Q^2)\pm \sqrt{(2M^2-Q^2)^2-4J^2}]}$\\

    $\sum T_{i}$:   &$\frac{a^2+Q^2-M^2}{2\pi M(a^2+Q^2)} $&$\frac{a^2-M^2}{2\pi aJ}$&$\frac{4J^{2}-(2M^2-Q^2)^{2}}{8\pi M J^{2}}$\\

    $\prod T_{i}$ : &$\frac{a^2+Q^2-M^2}{(4\pi M)^2(a^2+Q^2)}$ &$\frac{a^2-M^2}{(4\pi J)^2}$& $ \frac{4J^{2}-(2M^2-Q^2)^{2}}{(8\pi JM)^{2}}$\\

    $M_{irr, \pm}$: &$\sqrt{\frac{{\cal A}_{\pm}}{16\pi}}$&$\sqrt{\frac{{\cal A}_{\pm}}{16\pi}}$ &$\sqrt{\frac{{\cal A}_{\pm}}{16\pi}}$\\

    $\sum M_{irr,\pm}^{2}$: &$M^{2}$&$M^{2}$&$M^2-\frac{Q^{2}}{2}$\\

    $\prod M_{irr, \pm}$: &$\sqrt{\frac{J^2+\frac{Q^4}{4}}{4}}$&$\frac{J}{2}$ & $\frac{J}{2}$\\

    $\Omega_{\pm}$: &$\frac{a}{2Mr_{\pm}-Q^2}$ & $\frac{a}{2Mr_{\pm}}$ &
    $\frac{a}{2Mr_{\pm}}$ \\

    $\sum \Omega_{i} $: &$\frac{2a(2M^2-Q^2)}{4J^2+Q^{4}}$ & $\frac{1}{a}$ & $\frac{2M^2-Q^2}{2aM^2}$\\

    $\prod \Omega_{i}$ : &$\frac{a^{2}}{4J^2+Q^{4}}$ & $\frac{1}{4M^{2}}$&$\frac{1}{4M^{2}}$\\

    $E_{\pm}$ : &$\pm\sqrt{M^2-a^2-Q^2}$&$\pm\sqrt{M^2-a^2}$&
     $\pm \sqrt{(2M^2-Q^2)^2-4J^2}$\\

    $\sum E_{i} $: & $0$&$0$&$0$\\

    $\prod E_{i}$ : &$-(M^2-a^2-Q^2)$&$-(M^2-a^2)$&$-[(2M^2-Q^2)^2-4J^2]$\\

    $r_{+}=r_{-}$: & $ M^{2}=a^2+Q^{2}$ & $M^{2}=a^2$ & $a=M-\frac{Q^{2}}{2M}$ \\

    \hline
\end{tabular}
\end{center}

\section{Appendix}
Here we shall derive the central charges $c_{L}$ and $c_{R}$ of the left and right moving sectors of the 
dual CFT in Sen/CFT correspondence. We shall prove that the central charges of the left and right moving 
sectors are same i.e. $c_{L}=c_{R}$ for Sen BH. Also we shall derive the dimensionless  temperature of 
microscopic CFT from the above thermodynamic relations. Furthermore using Cardy formula,  we shall derive the left and right moving entropies in 2D CFT.

In terms of $r_{+}$ and $r_{-}$, we can write the ADM mass and spin parameter as 
\begin{eqnarray}
M=\frac{1}{4} \left[(r_{+}+ r_{-})+\sqrt{(r_{+}+ r_{-})^2+8Q^2}\right] \,\, \mbox{and}\,\, a=\sqrt{r_{+} r_{-}}  
~.\label{ss1}
\end{eqnarray}

Now the angular momentum can be written as 
\begin{eqnarray}
J=\frac{\sqrt{r_{+} r_{-}}}{4} \left[(r_{+}+ r_{-})+\sqrt{(r_{+}+ r_{-})^2+8Q^2}\right]   ~.\label{ss2}
\end{eqnarray}

Moreover using $r_{+}$ and $r_{-}$, we can write the entropy, Hawking temperature, angular velocity
and electric potential for  ${\cal H}^{+}$:
\begin{eqnarray}
S_{+} &=& \frac{\pi r_{+} }{2} \left[(r_{+}+ r_{-})+\sqrt{(r_{+}+ r_{-})^2+8Q^2}\right]   ~.\label{ss3}\\
T_{+} &=& \frac{ r_{+}-r_{-} }{2\pi r_{+} \left[(r_{+}+ r_{-})+\sqrt{(r_{+}+ r_{-})^2+8Q^2}\right]} ~.\label{ss4}\\
\Omega_{+} &=& \frac{2 \sqrt{r_{+} r_{-}}}{r_{+}\left[(r_{+}+ r_{-})+\sqrt{(r_{+}+ r_{-})^2+8Q^2}\right]} ~.\label{ss5}\\
\phi_{+} &=& \frac{4Q}{\left[(r_{+}+ r_{-})+\sqrt{(r_{+}+ r_{-})^2+8Q^2}\right]} ~.\label{ss6}
\end{eqnarray}
Finally, using the symmetry of $r_{\pm}$, one can obtain the following relations for the 
thermodynamic quantities at  ${\cal H}^{-}$:
\begin{eqnarray}
T_{-} &=& -T_{+}|_{r_{+}\leftrightarrow r_{-}}, S_{-}=S_{+}|_{r_{+}\leftrightarrow r_{-}}, 
\Omega_{-}=\Omega_{+}|_{r_{+}\leftrightarrow r_{-}}, \Phi_{-}=\Phi_{+}|_{r_{+}\leftrightarrow r_{-}} 
~.\label{ss7}
\end{eqnarray}
The first law of BH thermodynamics can be rewritten as in terms of left and right moving modes of dual 
CFT:
\begin{eqnarray}
\frac{dM}{2} &=& T_{L} dS_{L}+\Omega_{L} dJ+\Phi_{L} dQ   ~.\label{ss8} \\
             &=& T_{R} dS_{R}+\Omega_{R} dJ+\Phi_{R} dQ   ~.\label{ss9}
\end{eqnarray}
with the definitions $\beta_{R,L}=\beta_{+}\pm \beta_{-}$, $\beta_{\pm}=\frac{1}{T_{\pm}}$, 
$\Omega_{R,L}=\frac{\beta_{+}\Omega_{+}\pm \beta_{-}\Omega_{-}}{2 \beta{R,L}}$,  
$\Phi_{R,L}=\frac{\beta_{+}\Phi_{+}\pm \beta_{-}\Phi_{-}}{2\beta{R,L}}$ and 
$S_{R,L}=\frac{(S_{+}\mp S_{-})}{2}$. 

Using the above relations, we find
\begin{eqnarray}
T_{L} &=& \frac{1}{2\pi \left[(r_{+}+ r_{-})+\sqrt{(r_{+}+ r_{-})^2+8Q^2}\right]}, \,\,\, 
T_{R} = \frac{r_{+}-r_{-}}{2\pi (r_{+}+r_{-}) \left[(r_{+}+ r_{-})+\sqrt{(r_{+}+ r_{-})^2+8Q^2}\right]} \nonumber\\
S_{L} &=&  \frac{\pi (r_{+}+r_{-})}{4} \left[(r_{+}+ r_{-})+\sqrt{(r_{+}+ r_{-})^2+8Q^2}\right], \,\,\, 
S_{R} =  \frac{\pi (r_{+}-r_{-})}{4} \left[(r_{+}+ r_{-})+\sqrt{(r_{+}+ r_{-})^2+8Q^2}\right] \nonumber\\
\Omega_{L} &=& 0,  \,\,\, \Omega_{R} = \frac{2 \sqrt{r_{+} r_{-}}}{(r_{+}+r_{-})
\left[(r_{+}+ r_{-})+\sqrt{(r_{+}+ r_{-})^2+8Q^2}\right]} \nonumber\\
\Phi_{L} &=& \frac{2Q}{\left[(r_{+}+ r_{-})+\sqrt{(r_{+}+ r_{-})^2+8Q^2}\right]}, \,\,\, 
\Phi_{R} = \frac{2Q}{\left[(r_{-}+ r_{+})+\sqrt{(r_{-}+ r_{+})^2+8Q^2}\right]}  ~.\label{ss10}
\end{eqnarray}
Using Eqs. (\ref{ss8},\ref{ss9}) and setting $dQ=0$, we obtain the first law of left and right 
sectors:
\begin{eqnarray}
dJ &=& \frac{T_{L}}{\Omega_{R}-\Omega_{L}} dS_{L}- \frac{T_{R}}{\Omega_{R}-\Omega_{L}} dS_{R} ~.\label{ss11}
\end{eqnarray}
This gives the dimensionless temperature of the left and right moving sectors of the dual CFT correspondence
and are given by 
\begin{eqnarray}
T_{L,R}^{J} &=& \frac{T_{L,R}}{\Omega_{R}-\Omega_{L}}  ~.\label{ss12}
\end{eqnarray}
which is exactly the microscopic temperature of the CFT and found to be for Sen BH
\begin{eqnarray}
T_{L,R}^{J} &=& \frac{r_{+}\pm r_{-}}{4\pi \sqrt{r_{+} r_{-}}}  ~.\label{ss13}
\end{eqnarray}
Now we find the central charges\cite{chen12} in left and right moving sectors of the Sen/CFT 
correspondence via the Cardy formula reads
\begin{eqnarray}
S_{L,R}^{J} &=& \frac{\pi^2}{3}c_{L,R}^{J}T_{L,R}^{J}    ~.\label{ss14}
\end{eqnarray}
Therefore the central charges of dual CFT should be 
\begin{eqnarray}
c_{L}^{J} &=& c_{R}^{J}=12J    ~.\label{ss15}
\end{eqnarray}
which is exactly same as Kerr BH\cite{hartman9} and Kerr-Newman BH\cite{chen12}. This observation tells us 
that Sen BH is dual to a $c_{L}=c_{R}=12J$ 2D CFT at temperature $(T_{L},T_{R})$ for each value of $M$ and $J$.

In the extremal limit $r_{+}=r_{-}$, the above expressions reduce to 
\begin{eqnarray}
T_{L} &=& \frac{1}{4\pi \left[r_{+}+\sqrt{r_{+}^2+2Q^2}\right]}, \,\,\,  T_{R} = 0 \nonumber\\
S_{L} &=& \pi r_{+}\left[r_{+}+\sqrt{r_{+}^2+2Q^2}\right]  , \,\,\, 
S_{R} = 0 \nonumber\\
\Omega_{L} &=& 0,  \,\,\, \Omega_{R} = \frac{1}{2 \left[r_{+}+\sqrt{r_{+}^2+2Q^2}\right]} \nonumber\\
\Phi_{L} &=& \Phi_{R} = \frac{Q}{\left[r_{+}+\sqrt{r_{+}^2+2Q^2}\right]} ~.\label{ss16}
\end{eqnarray}
\begin{eqnarray}
T_{L}^{J} &=&  \frac{1}{2\pi}, \,\,\,  T_{R}^{J} = 0  ~.\label{ss17}
\end{eqnarray}
this left moving temperature is actually Frolov-Thorn temperature, and finally the central charge for 
extremal Sen BH:
\begin{eqnarray}
c_{L}^{J} &=& 12J ~.\label{ss18}
\end{eqnarray}
Therefore, we obtain the microscopic entropy via the Cardy formula in chiral dual CFT:
\begin{eqnarray}
S_{micro} &=& \frac{\pi^2}{3}c_{L}^{J}T_{L}^{J}=2\pi J    ~.\label{ss19}
\end{eqnarray}
which is perfectly agreement with macroscopic Bekenstein-Hawking entropy of the extreme Sen BH.

\section*{Acknowledgments}
I  would like to thank the Editor for his patience and anonymous referee for his helpful suggestions.


\begin{thebibliography}{99}

\bibitem{bk72} J. D. Bekenstein, \textit{Lett. Nuov. Cimento}  {\bf 4} {737} (1972).

\bibitem{bk73} J. D. Bekenstein, \textit{ Phys. Rev.} {\bf D 7} {2333} (1973).

\bibitem{bk74} J. D. Bekenstein, \textit{Phys. Rev.} {\bf D 9} {3292} (1974).

\bibitem{bcw73} J. M. Bardeen, B. Carter and S. W. Hawking, \textit{Commun. Math. Phys.} {\bf 31}, {161} (1973).

\bibitem{blaga01} P. A. Blaga and C. Blaga, \textit{Class. Quant. Grav.} {\bf 18}  3893-3905 (2001).

\bibitem{harm99} R. Casadio and B. Harms, \textit{Mod. Phys. Lett. A} {\bf 14} 1098 (1999).

\bibitem{smarr73} L. Smarr, \textit{ Phys. Rev. Lett.}  {\bf 30} 71 (1973), [Erratum-ibid. 30, 521 (1973)].


\bibitem{ansorg09} M. Ansorg and J. Hennig, \textit{ Phys. Rev. Lett.}  {\bf 102} {221102} (2009).


\bibitem{visser12} M. Visser, \textit{ J. High Energy Phys.}  {\bf 06} {023} (2012).

\bibitem{pp14} P. Pradhan,  \textit{The European Physical Journal C} {74} {2887} (2014).

\bibitem{hl15}  P. Pradhan, \textit{Phys. Lett. } {B 747} {64} (2015).

\bibitem{cr79} A. Curir, \textit{ Nuovo Cimento} {\bf 51B } {262} (1979).

\bibitem{mcdy96} M. Cveti\v{c} and D. Youm, \textit{ Phys. Rev. }{\bf D 54} {2612} (1996).

\bibitem{mcflb97} M. Cveti\v{c} and F. Larsen, \textit{ Nucl. Rhys.} {\bf B 506} {107} (1997).

\bibitem{mcfl97} M. Cveti\v{c} and F. Larsen, \textit{ Phys. Rev.} {\bf D 56} {4994} (1997).


\bibitem{mcgw11} M. Cveti\v{c}, G. W. Gibbons and C. N. Pope, \textit{ Phys. Rev. Lett.} {\bf 106} {121301} (2011).

\bibitem{pope14} M. Cveti\v{c}, H. L\"{u},  and C. N. Pope, \textit{Phys. Rev.} {\bf D 88} {044046} (2013).

\bibitem{castro12} A. Castro and M. J. Rodriguez, \textit{Phys. Rev.} {\bf D 86} {024008} (2012).

\bibitem{sia13}  A. M. Ghezelbash and H. M. Siahaan, \textit{Class. Quant. Grav.} {\bf 30} 135005 (2013).

\bibitem{det12} S. Detournay, \textit{Phys. Rev. Lett. } {\bf 109}, {031101}  (2012).

\bibitem{val13} V. Faraoni, A. F. Z. Moreno,  \textit {Phys. Rev. D. } {\bf 88} 044011 (2013).

\bibitem{chen12} B.~Chen, S.~X.~Liu and J.~J.~Zhang, \textit{ J. High Energy Phys.}  {\bf 017}, 1211 (2012).

\bibitem{sch} S. Chandrashekar, \textit{ The Mathematical Theory of Black Holes}, Clarendon Press, Oxford (1983).

\bibitem{as92} A. Sen, \textit{ Phys. Rev. Lett.} {\bf 69}, {1006} (1992).

\bibitem{as94} A. Sen, \textit{ Nucl. Phys.} {\bf B 440} {421} (1995).

\bibitem{komar59} A. Komar, \textit {Phys. Rev. } {\bf 113}, 934 (1959).

\bibitem{peni} R. Penrose, \textit{ Ann. N. Y. Acad. Sci. } {\bf 224} {125} (1973).

\bibitem{xu}  W.~Xu, J. ~Wang, and X. ~Meng, \textit{Phys. Lett. } {B 746} {53} (2015).

\bibitem{cr71} D. Christodoulou and R. Ruffini, \textit{ Phys. Rev. D.}  {\bf 4} {3552} (1971).

\bibitem{ramesh} R. Narayan, \textit{Astronomy and Geophysics} {\bf 44} (2003).  

\bibitem{kaburki91} O. Kaburki and I. Okamoto, \textit{ Phys. Rev. D.}  {\bf 43} {340} (1991).

\bibitem{kaburki92} I. Okamoto and O. Kaburki, \textit{ Mon. Not. R. astr. Soc.}  {\bf 255} {539} (1992).

\bibitem{frolov89} V. P. Frolov and I. D. Novikov, \textit{Black Hole Physics}, Kluwer Academic Press (1989).

\bibitem{hartman9} T. Hartman, K. Murata, T. Nishioka and A. Strominger, \textit{J. High Energy Phys.} {\bf 04}, 019 (2009).

\end{thebibliography}
\end{document}